\documentclass[a4paper,12pt]{article}
\usepackage{amssymb,amsfonts,amsmath,mathtext,cite,enumerate,float,color}
\usepackage{amssymb,amsthm,latexsym, braket, euscript,amscd, authblk}

\usepackage[russian,english]{babel}
\usepackage[utf8]{inputenc}

\title{On classical capacity of Weyl channels}

\author[1]{G.G.Amosov\thanks {This work is supported by the Russian Science Foundation under grant N 19-11-00086.}}

\affil[1]{Steklov Mathematical Institute of Russian Academy of Sciences}

\begin{document}

\maketitle

\begin{abstract} 

The additivity of minimal output entropy is proved for the Weyl channel obtained by the deformation of a q-c Weyl channel. The classical capacity of channel is calculated.

\end{abstract}


\section{Introduction} 

The quantum coding theorem proved independently by A.S. Holevo \cite{Holevo} and B. Schumacher, M.D. Westmoreland \cite {SW} posed the task of calculating the Holevo upper bound $\overline {C}(\Phi ^{\otimes N})$ for a tensor product of N copies of quantum channel $\Phi $ because a classical capacity of $\Phi $ is given by the formula
$$
C(\Phi )=\lim \limits _{N\to +\infty }\frac {\overline {C}(\Phi ^{\otimes N})}{N}.
$$
The additivity conjecture asks whether the equality
\begin{equation}\label{ap}
\overline {C}(\Phi \otimes \Omega )=\overline {C}(\Phi )+\overline {C}(\Omega )
\end{equation}
holds true for the fixed channel $\Phi $ and an arbitrary channel $\Omega $. If the additivity property (\ref {ap}) takes place for $\Phi $ the classical capacity can be calculated as follows
\begin{equation}\label{AD}
C(\Phi )=\overline {C}(\Phi ).
\end{equation}
The same level of interest has the additivity conjecture in the weak form asking whether
$$
\overline {C}(\Phi ^{\otimes N})=N\overline {C}(\Phi )
$$
takes place for a fixed channel $\Phi$. The validity of this statement also leads to (\ref {AD}).
The additivity conjecture for $\overline C$ is closely related to the additivity conjecture for the minimal output entropy of a channel and the multiplicativity conjectures for trace norms of a channel \cite {AH}. At the moment, the additivity is proved for many significant cases \cite {King1, King2,  Shor, Ho1, Ho2} including the solution to the famous problem of Gaussian optimizers \cite {Ho3, Ho4}. On the other hand, there are channels for which the additivity conjecture doesn't hold true \cite{Hast}. Recently the method of majorization was introduced to estimate the Holevo upper bound for Weyl channels \cite {urReh}. In the present paper we prove the additivity conjecture for one subclass of Weyl channels that are "deformations" of q-c channels of \cite {Holevo}. Our method is based upon \cite{urReh}.

Throughout this paper we denote $\mathfrak {S}(H)$ the set of positive unit-trace operators (quantum states) in a Hilbert space $H$, $I_{H}$ is the identity operator in $H$ and $S(\rho )=-Tr(\rho \log\rho )$ is the von Neumann entropy of $\rho \in \mathfrak {S}(H)$. Quantum channel $\Phi :\mathfrak {S}(H)\to \mathfrak {S}(K)$ is a completely positive trace preserving map between the algebras of all bounded operators $B(H)$ and $B(K)$  in Hilbert spaces $H$ and $K$ respectively.
Given two $\rho ,\sigma \in \mathfrak {S}(H)$ for which $supp\rho \subset supp\sigma $ the quantum relative entropy is $S(\rho\ ||\ \sigma )=Tr(\rho \log\rho) -Tr(\rho\log\sigma )$. The property of non-increasing the relative entropy with respect to the action of a quantum channel $\Phi $ states \cite {Lind}
$$
S(\Phi (\rho )\ ||\ \Phi (\sigma ))\le S(\rho\ ||\ \sigma )
$$
for $\rho ,\sigma \in \mathfrak {S}(H)$.

The Holevo upper bound for a quantum channel $\Phi $ is determined by the formula
$$
\overline C(\Phi )=\sup \limits _{\pi _j,\rho _j\in \mathfrak {S}(H)}(S(\sum \limits _j\pi _j\Phi (\rho _j))-\sum \limits _j\pi _jS(\Phi (\rho _j))),
$$
where the supremum is taken over all probability distributions $(\pi _j)$ on the ensemble of states $\rho _j\in \mathfrak {S}(H)$.

\section{Weyl channels}

Here we use the techniques introduced in \cite {Amo2, Amo} and developed in \cite{Amo3, Amo4, Amo5, Amo6}.
Fix an orthonormal basis $(e_j,\ j\in {\mathbb Z}_n)$ in a Hilbert space $H$ with dimension $dimH=n,$ and consider two unitary operators in $H$ defined by the formula
\begin{equation}\label{form}
Ue_j=e^{\frac {2\pi i}{n}j}e_j,\ Ve_j=e_{j+1},\ j\in {\mathbb Z}_n.
\end{equation}
Formula (\ref {form}) determines unitaries $W_{jk}=U^jV^k$ called Weyl operators satisfying the property
\begin{equation}\label{id}
\sum \limits _{j,k\in {\mathbb Z}_n}W_{jk}\rho W_{jk}^*=nI_H,\ \rho\in \mathfrak {S}(H).
\end{equation}
Quantum channels of the form
\begin{equation}\label{weyl}
\Phi (\rho )=\sum \limits _{j,k\in {\mathbb Z}_n}\pi _{jk}W_{jk}\rho W_{jk}^*,\ \rho \in \mathfrak {S}(H),
\end{equation}
where $(\pi _{jk})$ is a probability distribution,
are said to be Weyl channels.
Given a unitary representation $\lambda $ of ${\mathbb Z}_n$ in $H$ and a probability distribution $(p_k,\ k\in {\mathbb Z}_n)$ a Weyl channel of the form
$$
\Psi _{\lambda }(\rho )=\sum \limits _{k\in {\mathbb Z}_n}p_k\lambda (k)\rho \lambda (k)^*,\ \rho \in \mathfrak {S}(H),
$$
is said to be a phase damping channel.

Let us fix a phase damping channel of the form
$$
\Psi (\rho )=\sum \limits _{k\in {\mathbb Z}_n}p_kV^k\rho V^{k*},\ \rho \in \mathfrak {S}(H),
$$
where $p=(p_k,\ k\in {\mathbb Z}_n)$ is a probability distribution.
Consider the quantum channel
\begin{equation}\label{channel}
\Phi (\rho )=\frac {1}{n}\sum \limits _{j\in {\mathbb Z}_n}U^j\Psi (\rho )U^{j*}=\frac {1}{n}\sum \limits _{j,k\in {\mathbb Z}_n}p_kU^jV^k\rho V^{k*}U^{j*}, 
\rho \in \mathfrak {S}(H),
\end{equation}
Formula (\ref {channel}) gives a general form of the Weyl channel invariant with respect to the action of the group $(U^j,\ j\in {\mathbb Z}_n)$ in the sense
\begin{equation}\label{invar}
U^j\Phi (\rho)U^{j*}=\Phi (\rho ),\ \rho \in \mathfrak {S}(H),\ j\in {\mathbb Z}_n.
\end{equation}
It follows from (\ref {invar}) that
\begin{equation}\label{qc}
{\mathbb E}\circ \Phi =\Phi ,
\end{equation}
where the expectation $\mathbb E$ to the algebra of fixed elements with respect to the action of $(U^j,\ j\in {\mathbb Z}_n)$ is given by 
$$
{\mathbb E}(\rho )=\frac {1}{n}\sum \limits _{j\in {\mathbb Z}_n}U^j\rho U^{j*},\ \rho \in \mathfrak {S}(H).
$$

Put
$$
\Xi _k(\rho )=\frac {1}{n}\sum \limits _{j\in {\mathbb Z}_n}U^jV^k\rho V^{k*}U^{j*},\ \rho \in \mathfrak {S}(H),\ j\in {\mathbb Z}_n,
$$
then (\ref {channel}) can be represented as
$$
\Phi (\rho )=\sum \limits _{k\in {\mathbb Z}_n}p_k\Xi _k(\rho ).
$$

The property (\ref {qc}) shows that $\Phi $ is a q-c channel and the additivity of $\overline C$ was shown in \cite {Holevo}. We place the following statement here to calculate the exact value of a classical capacity.

{\bf Proposition 1.} {\it Given a quantum channel $\Omega :\mathfrak {S}(K)\to \mathfrak {S}(K)$ and a pure state $\ket {\xi }\bra {\xi }\in \mathfrak {S}(H\otimes K)$
$$
\inf \limits _{\rho \in \mathfrak {S}(H\otimes K)}S(\Phi \otimes \Omega (\ket {\xi }\bra {\xi } ))\ge -\sum \limits _{k\in {\mathbb Z}_n}p_k\log p_k+S(\Omega (Tr_H(\ket {\xi }\bra {\xi } ))).
$$
}

Proof.

Let us define a c-q channel $\Upsilon :\mathfrak {S}(H)\to \mathfrak {S}(H\otimes K)$ by the formula
$$
\Upsilon (\rho )=\sum \limits _{k\in {\mathbb Z}_n}\braket {e_k,\rho e_k}(\Xi _k\otimes \Omega )(\ket {\xi }\bra {\xi }),\ \rho \in \mathfrak {S}(H).
$$
Put
\begin{equation}\label{1}
\rho =\sum \limits _{k\in {\mathbb Z}_n}p_k\ket {e_k}\bra {e_k},\ \sigma =\frac {1}{n}I_H.
\end{equation}
Applying the property of non-increasing the quantum relative entropy with respect to the action of quantum channel we obtain
\begin{equation}\label{3}
S(\Upsilon (\rho )\ ||\ \Upsilon (\sigma ))\le S(\rho\ ||\ \sigma ).
\end{equation}
It follows from (\ref {id}) that 
$$
\sum \limits _{k\in {\mathbb Z}_n}\Xi _k(\rho )=I_H,\ \rho \in \mathfrak {S}(H).
$$
Hence
$$
\sum \limits _{k\in {\mathbb Z}_n}(\Xi _k\otimes \Omega )(\ket {\xi }\bra {\xi })=I_H\otimes \Omega (Tr_H(\ket {\xi }\bra {\xi }))
$$
and
\begin{equation}\label {2}
\Upsilon (\sigma )=\frac {1}{n}I_H\otimes \Omega (Tr_H(\ket {\xi }\bra {\xi })).
\end{equation}
Substituting (\ref {1})--(\ref {2}) to (\ref {3}) we get
$$
-S((\Phi \otimes \Omega )(\ket {\xi }\bra {\xi }))-Tr\left ((\Phi \otimes \Omega )(\ket {\xi }\bra {\xi })\log\left (\frac {1}{n}I_H\otimes \Omega (Tr_H(\ket {\xi }\bra {\xi}))\right )\right )\le 
$$
$$
\sum \limits _{k\in {\mathbb Z}_n}p_k\log p_k-Tr(\rho \log \sigma ).
$$ 
Taking into account that
$$
Tr\left ((\Phi \otimes \Omega )(\ket {\xi }\bra {\xi })\log\left (\frac {1}{n}I_H\otimes \Omega (Tr_H(\ket {\xi }\bra {\xi}))\right )\right )=-\log n-S(\Omega (Tr_H(\ket {\xi }\bra {\xi })))
$$
and
$$
Tr(\rho \log \sigma )=-\log n
$$
we obtain the result.

$\Box $

{\bf Corollary 1.} {\it Given a quantum channel $\Omega :\mathfrak {S}(K)\to \mathfrak {S}(K)$ and the q-c Weyl channel  (\ref {channel}) the following equality holds
$$
\inf \limits _{\rho \in \mathfrak {S}(H\otimes K)}S((\Phi \otimes \Omega )(\rho ))=\inf \limits _{\rho \in \mathfrak {S}(H)}S(\Phi (\rho ))+\inf \limits _{\rho \in \mathfrak {S}(K)}S(\Omega (\rho )).
$$
}

Proof.

Notice that 
$$
S(\Phi (\ket {e_j}\bra {e_j}))=-\sum \limits _{k\in {\mathbb Z}_n}p_k\log p_k\ge \inf \limits _{\rho \in \mathfrak {S}(H)}S(\Phi (\rho ))
$$
for any $j\in {\mathbb Z}_n$. It follows from Proposition 1 that
\begin{equation}\label{rhs}
\inf \limits _{\rho \in \mathfrak {S}(H\otimes K)}S((\Phi \otimes \Omega )(\rho ))\ge \inf \limits _{\rho \in \mathfrak {S}(H)}S(\Phi (\rho ))+\inf \limits _{\rho \in \mathfrak {S}(K)}S(\Omega (\rho )).
\end{equation}
On the other hand, the right side in (\ref {rhs}) can not be less than the left hand side. Hence,
$$
\inf \limits _{\rho \in \mathfrak {S}(H)}S(\Phi (\rho ))=-\sum \limits _{k\in {\mathbb Z}_n}p_k\log p_k
$$
and we have the equality in (\ref {rhs}).

$\Box $

{\bf Corollary 2.}  {\it The classical capacity of the q-c Weyl channel  (\ref {channel}) is given by the formula
$$
C(\Phi )=\log(n)+\sum \limits _{k\in {\mathbb Z}_n}p_k\log p_k.
$$
}

Proof.

The statement can be derived from the fact that
$$
\overline C(\Phi ^{\otimes N})=N\log n-\inf \limits _{\rho \in \mathfrak {S}(H^{\otimes N})}S(\Phi ^{\otimes N}(\rho ))
$$
for covariant channels \cite {Hol}. It follows from Corollary 1 that
$$
\inf \limits _{\rho \in \mathfrak {S}(H^{\otimes N})}S(\Phi ^{\otimes N}(\rho ))=N\inf \limits _{\rho \in \mathfrak {S}(H)}S(\Phi (\rho )).
$$
In the proof of Corollary 1 we have shown that
\begin{equation}\label {e}
\inf \limits _{\rho \in \mathfrak {S}(H)}S(\Phi (\rho ))= -\sum \limits _{k\in {\mathbb Z}_n}p_k\log p_k.
\end{equation}

$\Box $

\section{Majorization}

Let $\mathfrak J$ be the index set and $|\mathfrak {J}|=d<+\infty $.
Given a probability distribution $\lambda =(\lambda _J,\ J\in \mathfrak {J})$ we denote $\lambda ^{\downarrow}=(\lambda ^{\downarrow}_j,\ 1\le j\le d)$ the probability distribution obtained by sorting $\lambda $ in the decreasing order,
$$
\lambda _1^{\downarrow}\ge \lambda _2^{\downarrow}\ge \dots \ge \lambda _d^{\downarrow}.
$$
Consider two probability distribution $\lambda =(\lambda _J,\ J\in \mathfrak {J})$ and $\mu =(\mu _J,\ J\in \mathfrak {J})$.   We shall say that $\lambda $ majorizes $\mu $ and write
$$
\mu \prec \lambda 
$$
iff
$$
\sum \limits _{j=1}^k\mu_j^{\downarrow} \le \sum \limits _{j=1}^k\lambda_j^{\downarrow} ,\ 1\le k\le d.
$$
Let $H_d$ be a Hilbert space with $dimH_d=d$. Denote $B(H_d)$ the algebra of all bounded operators in $H_d$.
The following statement can be derived from \cite{urReh} (see Theorem 2). 

{\bf Proposition 2.} {\it Let $0\le X_J\le I,\ J\in \mathfrak {J},\ |\mathfrak {J}|=d^2,$ be a set of positive operators in $B(H_d)$ such that
$$
\sum \limits _{J\in \mathfrak {J}}X_J=dI_{H_d}.
$$
Then, given a probability distribution $\pi =(\pi _J,\ J\in \mathfrak {J})$ 
the eigenvalues $\lambda =(\lambda _j)_{j=1}^d$  of the positive operator
$$
A=\sum \limits _{J\in \mathfrak {J}}\pi _JX_J
$$
sorted in the decreasing order $\lambda \equiv \lambda ^{\downarrow}$
satisfy the relation
$$
\lambda \prec p,
$$
where
$$
p_j=\sum \limits _{m=1+(j-1)d}^{d+(j-1)d}\pi _m^{\downarrow},\ 1\le j\le d. 
$$
}

Proof.

Let $(e_j)_{j=1}^d$ be the unit eigenvectors corresponding to the eigenvalues $(\lambda _j)_{j=1}^d$.
Then,
$$
\sum \limits _{j=1}^k\lambda _j=\sum \limits _{j=1}^k\braket {e_j,Ae_j}=\sum \limits _{j=1}^k\sum \limits _{J\in \mathfrak {J}}\pi _J\braket {e_j,X_Je_j}\le \sum \limits _{j=1}^kp_j,\ 1\le k\le d.
$$

$\Box $

{\bf Corollary 3.} {\it The eigenvalues $\lambda $ of the positive operator $A$ in Proposition 2 possess  the property
$$
-\sum \limits _{j=1}^d\lambda _j\log \lambda _j\ge -\sum \limits _{j=1}^dp_j\log p_j.
$$
}

Proof.

Since $\lambda $ majorizes $\mu $ due to Proposition 2, we get the result \cite {major}.

$\Box $

\section{Deformation of q-c Weyl channels}

Let us come back to Weyl channels (\ref {weyl}).

{\bf Definition.} {\it Suppose that
a probability distribution $(\pi_{jk},\ j,k\in {\mathbb Z}_n)$ satisfies the relation
\begin{equation}\label{uslovie}
\pi_{00}\ge \pi_{10}\ge \dots \ge \pi_{n-10}\ge \pi_{01}\ge \dots \pi_{n-11}\ge \pi_{02}\ge \dots \ge \pi_{n-1n-1}.
\end{equation}
Put
\begin{equation}\label{mrak}
p_k=\sum \limits _{j\in {\mathbb Z}_n}\pi _{jk},\ k\in {\mathbb Z}_n.
\end{equation}
Then (\ref {weyl})
is said to be the Weyl channel obtained by the deformation of q-c channel (\ref {channel}).
}

{\bf Theorem.} {\it 
The Weyl channel $\Phi $ obtained by the deformation of q-c channel satisfies the property
$$
\inf \limits _{\rho \in {\mathfrak S}(H^{\otimes N})}S(\Phi ^{\otimes N}(\rho ))=-N\sum \limits _{k=1}^{n}p_j\log p_j.
$$
}

Proof.

Denote $\mathfrak {J}$ the index set $({\mathbb Z}_n\times {\mathbb Z}_n)^{\times N}$ consisting of collections $(j_1,k_1),\dots ,(j_N,k_N),$ $ j_s,k_s\in {\mathbb Z}_n$. Let us consider the probability distribution $\Pi =(\Pi _J,\ J\in \mathfrak {J})$ and a set of positive operators $(X_J,\ J\in \mathfrak {J})$ defined by the formula
$$
\Pi _J=\prod _{s=1}^N\pi _{j_sk_s},
$$
$$
X_J=\left (\otimes _{s=1}^NW_{j_sk_s}\right )\rho \left (\otimes _{s=1}^NW_{j_sk_s}^*\right ),\ J\in \mathfrak {J},
$$
where $\rho $ is a fixed state in $\mathfrak {S}(H^{\otimes N})$.
Then, the conditions of Proposition 2 is satisfied for $(\Pi _J)$, $(X_J)$ and $d=n^N$. Applying Corollary 3 we obtain
\begin{equation}\label{fin}
S(\Phi (\rho ))\ge -N\sum \limits _{j=1}^Np_j\log p_j.
\end{equation}
The equality in (\ref {fin}) is achieved for any
$$
\rho =\ket {e}\bra {e},
$$
where
$$
e=\otimes _{s=1}^Ne_{j_s},\ j_s\in {\mathbb Z}_n.
$$
$\Box $

{\bf Corollary 4.}  {\it The classical capacity of the Weyl channel obtained by the deformation of (\ref {channel}) is given by the formula
$$
C(\Phi )=\log(n)+\sum \limits _{k\in {\mathbb Z}_n}p_k\log p_k.
$$
}

Proof.

The statement can be derived from the fact that
$$
\overline C(\Phi ^{\otimes N})=N\log n-\inf \limits _{\rho \in \mathfrak {S}(H^{\otimes N})}S(\Phi ^{\otimes N}(\rho ))
$$
for covariant channels \cite {Hol}. It follows from Theorem that
$$
\inf \limits _{\rho \in \mathfrak {S}(H^{\otimes N})}S(\Phi ^{\otimes N}(\rho ))=N\inf \limits _{\rho \in \mathfrak {S}(H)}S(\Phi (\rho ))=-N\sum \limits _{j=1}^Np_j\log p_j.
$$

$\Box $

\subsection{Example: qutrits}

Because the qubit case $dimH=2$ is completely parsed \cite {King1} a simplest example of the introduced techniques can be given for qutrits, $dimH=3$. Let us define two unitary operators $U$ and $V$ satisfying (\ref {form})
$$
Ue_0=e_0,\ Ue_1=e^{i\frac {2\pi }{3}}e_1,\ Ue_2=e^{i\frac {4\pi }{3}}e_2, 
$$
$$
Ve_0=e_1,\ Ve_1=e_2,\ Ve_2=e_0.
$$
Then, consider the expectation (\ref {qc})
$$
{\mathbb E}(x)=\frac {1}{3}\sum \limits _{j=0}^2U^jxU^{j*},\ x\in B(H).
$$
Taking a probability distribution $\{p_0,p_1,p_2\}$ we can define a qc Weyl channel by the formula
\begin{equation}\label{example}
\Phi _{qc}(\rho )={\mathbb E}\circ \sum \limits _{k=0}^2p_kV^k\rho V^{k*},\ \rho\in {\mathfrak S}(H).
\end{equation}
It follows from Corollary 1 and Corollary 2 that
$$
\inf \limits _{\rho \in \mathfrak {S}(H\otimes K)}S((\Phi _{qc}\otimes \Omega )(\rho ))=\inf \limits _{\rho \in \mathfrak {S}(H)}S(\Phi _{qc}(\rho ))+
\inf \limits _{\rho \in \mathfrak {S}(K)}S(\Omega (\rho ))
$$
for any quantum channel $\Omega :\mathfrak {S}(K)\to \mathfrak {S}(K)$ and
the classical capacity is equal to
$$
C(\Phi _{qc})=n+\sum \limits _{k=0}^2p_k\log p_k.
$$
Suppose that $p_0\ge p_1\ge p_2$ and one can pick up positive numbers $\pi _{jk},\ 0\le j,k\le 2,$ satisfying the relations
$$
\pi _{00}\ge \pi _{10}\ge \pi _{20}\ge \pi _{01}\ge \pi _{11}\ge \pi _{21}\ge \pi _{02}\ge \pi _{12}\ge \pi _{22},
$$
$$
p_k=\pi _{0k}+\pi _{1k}+\pi _{2k},\ 0\le k\le 2.
$$
Then,
$$
\Phi (\rho )=\sum \limits _{j,k=0}^2\pi _{jk}U^jV^k\rho V^{k*}U^{j*},\ \rho \in \mathfrak {S}(H),
$$
is the Weyl channel obtained by the deformation of (\ref {example}). Applying Corollary 4 we obtain for a classical capacity
$$
C(\Phi )=\log (3)+p_0\log p_0+p_1\log p_1+ p_2\log p_2.
$$
As a concrete example one can take
$$
p_0=\frac {1}{2},\ p_1=\frac {1}{3},\ p_2=\frac {1}{6}.
$$
In the case, one of possible deformations is given by
$$
\pi _{00}=\frac {1}{4},\ \pi _{10}=\frac {1}{8},\ \pi _{20}=\frac {1}{8},
$$
$$
\pi _{01}=\frac {1}{8},\ \pi _{11}=\frac {1}{8},\ \pi _{21}=\frac {1}{12},
$$
$$
\pi _{02}=\frac {1}{12},\ \pi _{12}=\frac {1}{24},\ \pi _{22}=\frac {1}{24}.
$$

\section*{Acknowledgments} The author is grateful to A.S. Holevo for fruitful discussion and useful comments.

\end{document}